\documentclass{article}
\PassOptionsToPackage{numbers, compress}{natbib}
\usepackage{natbib}
\usepackage[final]{main}
\usepackage[utf8]{inputenc} % allow utf-8 input
\usepackage[T1]{fontenc}    % use 8-bit T1 fonts
\usepackage[hidelinks]{hyperref}       % hyperlinks
\usepackage{url}            % simple URL typesetting
\usepackage{booktabs}       % professional-quality tables
\usepackage{amsfonts}       % blackboard math symbols
\usepackage{nicefrac}       % compact symbols for 1/2, etc.
\usepackage{microtype}      % microtypography
\usepackage{xcolor}         % colors
\usepackage{caption}
\usepackage{subcaption}
\usepackage{array, longtable}
\usepackage{amssymb}
\usepackage{graphicx}
\usepackage{pifont}
\newcommand{\cmark}{\ding{51}}
\newcommand{\xmark}{\ding{55}}

\usepackage{tabularx}
\usepackage{amsmath}
\newcolumntype{C}{>{\centering\arraybackslash}X}
\title{Natural Language Generation and Understanding of Big Code for AI-Assisted Programming: A Review}

\author{%
  Man Fai Wong \\
  City University of Hong Kong\\
  \texttt{mfwong29-c@my.cityu.edu.hk} \\
   \And
  Shangxin Guo\\
  City University of Hong Kong\\
  \texttt{sxguo2-c@my.cityu.edu.hk} \\
  \And
  Ching Nam Hang\\
  City University of Hong Kong\\
  \texttt{cnhang3-c@my.cityu.edu.hk} \\
  \And
  Siu Wai Ho\\
  University of Adelaide\\
  \texttt{siuwai.ho@adelaide.edu.au} \\
  \And
  Chee Wei Tan\\
  Nanyang Technological University\\
  \texttt{cheewei.tan@ntu.edu.sg} \\
}

\begin{document}

\maketitle

\begin{abstract}
  This paper provides a comprehensive review of the literature concerning the utilization of Natural Language Processing (NLP) techniques, with a particular focus on transformer-based large language models (LLMs) trained using Big Code, within the domain of AI-assisted programming tasks. LLMs, augmented with software naturalness, have played a crucial role in facilitating AI-assisted programming applications, including code generation, code completion, code translation, code refinement, code summarization, defect detection, and clone detection. Notable examples of such applications include the GitHub Copilot powered by OpenAI's Codex and DeepMind AlphaCode. This paper presents an overview of the major LLMs and their applications in downstream tasks related to AI-assisted programming. Furthermore, it explores the challenges and opportunities associated with incorporating NLP techniques with software naturalness in these applications, with a discussion on extending AI-assisted programming capabilities to Apple's Xcode for mobile software development. This paper also presents the challenges of and opportunities for incorporating NLP~techniques with software naturalness, empowering developers with advanced coding assistance and streamlining the software development process.
\end{abstract}

\section{Introduction}
The advent of Big Code has become increasingly relevant in today's software development landscape as the size and complexity of software systems continue to grow~\cite{vechev2016programming}. Big Code refers to the vast collection of online software artifacts such as source code repositories, bug databases, and~code snippets. It represents a wealth of knowledge and experience that researchers can draw upon to improve the quality and efficiency of their own projects. The~goal of Big Code is to build tools and techniques that can assist software engineers to analyze, understand, and~make predictions about large codebases in a scalable and efficient manner. Big Code also has the potential to revolutionize artificial intelligence~(AI) development by unitizing  Big Code data. The~development of statistical programming systems involves the utilization of advanced programming languages, powerful machine learning techniques such as large language models~(LLMs), and~natural language processing~(NLP) techniques based on the software naturalness hypothesis~\cite{hindle2012naturalness}. This hypothesis posits that computer programs written in diverse programming languages can be comprehended and manipulated similarly to NLP's treatment of human natural languages. 

By~employing this combination of tools, probabilistic models of extensive codebases can be constructed. These systems query a probabilistic model and calculate the most probable predictions to solve a specific challenge \cite{goodman2001bit}, which are then presented to the developer. In~other words, the~programming language is regarded as the natural language for the NLP techniques in this study. There are several crucial areas of fundamental research focused on advancing probabilistic models of ``Big Code'' using statistical and machine learning methodologies. By~considering source code as a series of tokens and leveraging the inherent patterns and structures within vast code repositories, NLP techniques can be developed to enhance AI-assisted programming tasks, including code generation, code completion, code refinement, code summarization, defect detection, and~clone~detection. 

AI-assisted programming can enable software engineers to work more efficiently and effectively~\cite{dijkstra}, especially in~situations where complex algorithms are being used that involve large amounts of code (i.e., Big Code regime). It also strikes a balance between productivity and ensuring safety, security, and~reliability within the programming development environment~\cite{rajamani2022ai}. In~fact, this can even lead to the development of AI-based predictive analysis that allows human developers to more easily interact with code using natural language commands and queries as part of the software development process~\cite{dijkstra1972humble}. AI-based predictive analysis~\cite{ji2020amazing} can also more accurately anticipate potential issues throughout the software development life cycle and flag critical incidents~\cite{surameery2023use} before they occur~\cite{talamadupula2021applied,ross2023programmer}.

Several recent reviews have explored specific topics related to LLMs, such as fairness and bias~\cite{mehrabi2021survey}, interpretability~\cite{carvalho2019machine}, explainability~\cite{tjoa2020survey}, and~privacy preservation~\cite{beigi2020survey}. However, this review focuses primarily on language models with software naturalness. In~Table~\ref{tab:comparsion}, a~detailed comparison of other reviews that have examined related topics is provided. This review also delves into the analysis of the publicly available Big Code dataset, which is designed to assist programming with AI. This review addresses the process of using language models for assessing software naturalness and examines the concept of evaluating language models using entropy. Additionally, the~latest developments in AI-assisted programming using transformer-based LLMs trained on Big Code are explored, and~both the generation and comprehension aspects are discussed. The~review concludes with the open challenges and opportunities in AI-assisted programming. This review paper highlights the unique contributions of this review in comparison to existing~reviews.

Reviews have emphasized the significance of AI-assisted programming, leading to significant advancements in this critical field of study. However, the~essential components of AI-assisted programming have been presented separately, resulting in a fragmented understanding of the topic. Despite this, these independent studies have created an opportunity to view AI-assisted programming from a more comprehensive perspective. In~light of this, our survey aims to provide a more structured approach to framing \mbox{AI-assisted} programming that extends beyond the examination of individual research topics. By~doing so, this review paper hopes to offer a more comprehensive understanding of this field, highlighting the interdependencies between different areas of~research.

\begin{table}[h]
 \caption{Comparison of surveys on language models in software naturalness \label{tab:comparsion}}
		\begin{tabularx}{\textwidth}{p{8.5cm}<{\raggedright }p{0.7cm}<{\raggedright }p{3.5cm}}
			\toprule
			{\textbf{Title}} & {\textbf{Year}} & {\textbf{Focus Area}} \\
			\midrule
        A Survey of Machine Learning for Big Code and Naturalness~\cite{allamanis2018survey} &  2019  & Big Code and~Naturalness	 \\
        Software Vulnerability Detection Using Deep Neural Networks: A Survey~\cite{lin2020software} &  2020 & Security \\
        A Survey on Machine Learning Techniques for Source Code Analysis~\cite{sharma2021survey} & 2021 & Code~Analysis			 \\
        Deep Security Analysis of Program Code: A Systematic Literature Review~\cite{sonnekalb2022deep}	& 2022 & Security		 \\
        A Survey on Pretrained Language Models for Neural Code Intelligence~\cite{xu2022survey} & 2022 & Code Summarization and Generation, and~Translation \\
        Deep Learning Meets Software Engineering: A Survey on Pre-trained Models of Source Code~\cite{niu2022deep} &2022 & Software~Engineering\\
        Software as Storytelling: A Systematic Literature Review~\cite{ciancarini2023software} & 2023 & Storytelling  \\
        Pre-train, Prompt, and~Predict: A Systematic Survey of Prompting Methods in Natural Language Processing~\cite{liu2023pre}& 2023 & Prompt-based~Learning\\
\bottomrule
		\end{tabularx}
\end{table}

The remainder of this review article is structured as follows. Section~\ref{sec:bg} provides an overview of the background knowledge in Big Code and software naturalness, covering topics such as the available dataset, tokenization process, existing language models, and~the measurement of language models using entropy. Section~\ref{sec:task} explores recent applications of LLMs trained with Big Code in AI-assisted programming tasks. Section~\ref{sec:chop} discusses the potential challenges and opportunities associated with LLMs in this context. Finally, Section~\ref{sec:conc} concludes the study and outlines possible directions for future work in this~field.

\newpage
\section{Background}\label{sec:bg}
\unskip
\subsection{Main Big Code~Dataset}\label{sec:bcd}
Researchers have successively released a large amount of Big Code to train LLMs. Most datasets used to train LLMs can be applied into different tasks such as code generation and code summarization. LLMs use unsupervised learning and require large amounts of high-quality and diverse data to achieve high accuracy and generalization in their predictions. Access to large-scale, high-quality, diverse, and~representative datasets is essential for developing high-performing LLMs on software naturalness. The~datasets found in the literature are described in Table~\ref{tab:dataset}.

\begin{table}[!thb]
    \caption{Summary of public datasets used on Big Code}\label{tab:dataset}
	\begin{tabularx}{\textwidth}{p{2cm}<{\raggedright}p{0.5cm}<{\raggedright}p{1cm}<{\raggedright}p{1.8cm}<{\raggedright}p{3cm}<{\raggedright}p{3.3cm}<{\raggedright}}
			\toprule
	{\textbf{Dataset Name}} & {\textbf{Year}} & {\textbf{Sample Size}} & {\textbf{Language(s)}}& {\textbf{Supported Task(s)}}& {\textbf{Online URL}}\\
			\midrule
 GitHub Java Corpus~\cite{githubCorpus2013} &  2013 & 14.7K & Java & Code Completion & \url{https://groups.inf.ed.ac.uk/cup/javaGithub/}\\
    
    Description2- Code~\cite{Caballero2016} & 2016& 7.6K & Java, C\# & Code Generation, Code Summarization & \url{https://github.com/ethancaballero/description2code} \\
   
    BigClone- Bench~\cite{svajlenko2021bigclonebench} & 2015& 5.5K & Java & Defect Detection, Clone Detection & \url{https://github.com/clonebench/BigCloneBench} \\
    \bottomrule
		\end{tabularx}
\end{table}
\begin{table}[!thb]\ContinuedFloat
    \caption{Cont.}
	\begin{tabularx}{\textwidth}{p{2cm}<{\raggedright}p{0.5cm}<{\raggedright}p{1cm}<{\raggedright}p{1.8cm}<{\raggedright}p{3cm}<{\raggedright}p{3.3cm}<{\raggedright}}
			\toprule
	{\textbf{Dataset Name}} & {\textbf{Year}} & {\textbf{Sample Size}} & {\textbf{Language(s)}}& {\textbf{Supported Task(s)}}& {\textbf{Online URL}}\\
			\midrule
    CodRep~\cite{CodRep2018} & 2018 & 58K & Java &Code Refinement, Defect Detection & \url{https://github.com/ASSERT-KTH/CodRep}\\
    CONCODE~\cite{iyer2018mapping} & 2018 & 104K & Java & Code Generation & \url{https://github.com/sriniiyer/concode}\\
    WikiSQL~\cite{zhongSeq2SQL2017} & 2018 & 87K & SQL & Code Summarization& \url{https://github.com/salesforce/WikiSQL}\\
    Bugs2Fix~\cite{tufano2019empirical} & 2019 & 122K & Java & Defect Detection, Code Refinement & \url{https://sites.google.com/view/learning-fixes} \\
    Devign~\cite{zhou2019devign} & 2019 & 26.4K & C & Code Generation, Defect Detection & \url{https://sites.google.com/view/devign} \\
    CodeSearch- Net~\cite{husain2019codesearchnet} &2019 & 2M &  Python, Javascript, Ruby, Go, Java, PHP & Code Generation, Code Summarization, Code Translation & \url{https://github.com/github/CodeSearchNet}\\
        The Pile~\cite{pile} & 2020 & 211M& Python & Coder Generation & \url{https://pile.eleuther.ai}\\
    CodeNet~\cite{puri2021codenet} & 2021 & 13M&C++, C, Python, Java& Code Generation, Code Refinement &\url{https://github.com/IBM/Project_CodeNet}\\
    CodeX- GLUE~\cite{lu1codexglue} & 2021 & 176K & Python, Java, PHP, JavaScript, Ruby, Go &Code Generation, Code Completion, Code Summarization, Defect Detection & \url{https://github.com/microsoft/CodeXGLUE} \\
    HumanEval~\cite{chen2021evaluating}& 2021 &  164 & Python& Code Generation &\url{https://github.com/openai/human-eval}\\
    APPS~\cite{hendrycksapps2021} & 2021 &  10K & Python & Code Generation& \url{https://github.com/hendrycks/apps}\\
    Codeparrot~\cite{tunstall2022natural} &2022 & 22M &Python & Code Generation &\url{https://hf.co/datasets/transformersbook/codeparrot}\\
    Code- Contests~\cite{li2022competition} & 2022 & 13.6K & C++, Java, JavaScript, C\# and 8 more & Code Generation &\url{https://github.com/deepmind/code_contests}\\
    CERT~\cite{CERT} & 2022 & 5.4M & Python & Code Generation & \url{https://github.com/microsoft/PyCodeGPT}\\ 
    \bottomrule
		\end{tabularx}
\end{table}
\begin{table}[!thb]\ContinuedFloat
    \caption{Cont.}
	\begin{tabularx}{\textwidth}{p{2cm}<{\raggedright}p{0.5cm}<{\raggedright}p{1cm}<{\raggedright}p{1.8cm}<{\raggedright}p{3cm}<{\raggedright}p{3.3cm}<{\raggedright}}
			\toprule
	{\textbf{Dataset Name}} & {\textbf{Year}} & {\textbf{Sample Size}} & {\textbf{Language(s)}}& {\textbf{Supported Task(s)}}& {\textbf{Online URL}}\\
			\midrule
    InCoder~\cite{fried2022incoder} & 2022 & 670K & Python, JavaScript, HTML and 24 more & Code Generation, Code Summarization & \url{https://github.com/dpfried/incoder} \\
    PolyCoder~\cite{xu2022systematic} & 2022 & 1K & C, C++, Java, JavaScript, C\#, Go and 6 more & Code Generation & \url{https://github.com/VHellendoorn/Code-LMs} \\    
    ExecEval~\cite{khan2023xcodeeval} & 2023 & 58K & Ruby, Javascript, Go, C++, C and 6 more &Code Sumarization, Code Generation, Code Translation & \url{https://github.com/ntunlp/xCodeEval}\\
    \bottomrule
		\end{tabularx}
\end{table}
\subsection{Tokenization}\label{sec:token}

\begin{figure}[h]
\centering
\includegraphics[width=13.8 cm]{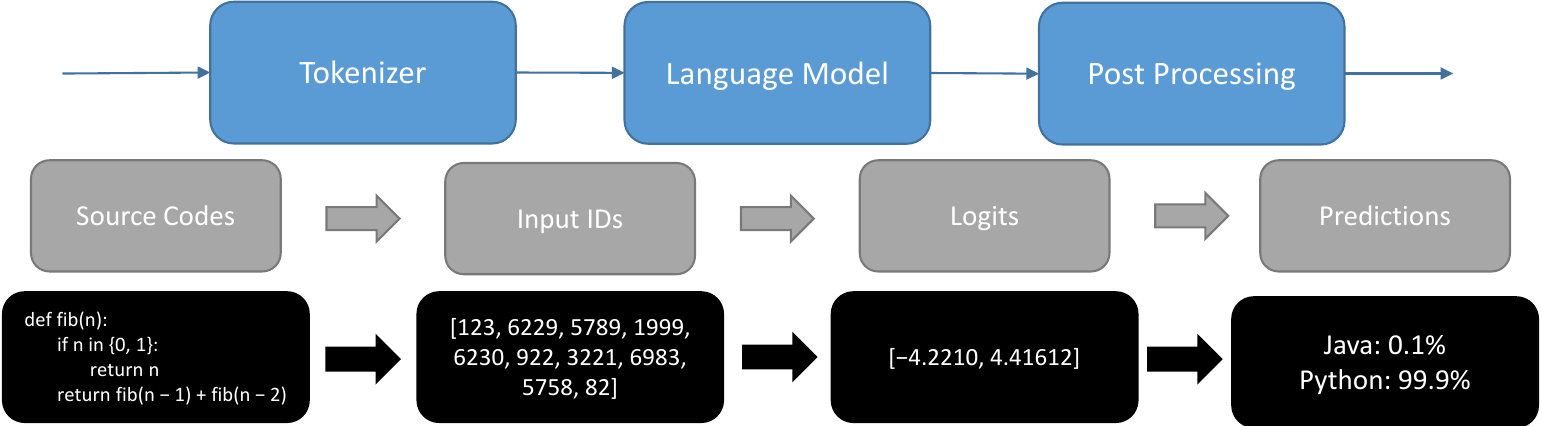}
\caption{Pipeline of language models on software naturalness.}\label{fig:pipeline}
\end{figure}

Figure \ref{fig:pipeline} illustrates the pipeline of language models on software naturalness. Similar to other neural networks and raw text, language models cannot process source code directly, so the first step of the standard pipeline is to convert the code inputs into numbers of which the model can make sense. To do this, a~tokenizer can be used to split the input into code syntax keyword, variables, or~symbols (similar to punctuation) that are called tokens. Each token is mapped to an integer in the next step. These tokens typically correspond to words, punctuation marks, or~other meaningful elements of the text. Tokenization is an important step in many NLP tasks, as~it allows machine learning algorithms to process and analyze text in a more efficient and meaningful way. Some popular tokenizers are available to be used directly such as Byte-Pair Encoding (BPE)~\cite{sennrichetal2016neural} and RoBERTa~\cite{liu2019roberta}.

In the tokenization process, each token is assigned a unique identifier or index which can be used to represent the token in a numerical format that can be understood by machine learning models. Different tokenization strategies may be used depending on the specific task at hand, such as splitting text into words, phrases, or~even individual characters. One common challenge in tokenization is dealing with ambiguity or variability in the text. For~example, words may have different meanings depending on the context in which they appear, or~may be misspelled or abbreviated in unpredictable ways. There are various techniques that can be used to address these challenges, such as using contextual information or statistical models to help disambiguate the~text.

\subsection{Language Models on Software~Naturalness}\label{sec:llm}
\begin{figure}[h]
\centering
\includegraphics[width=13.8 cm]{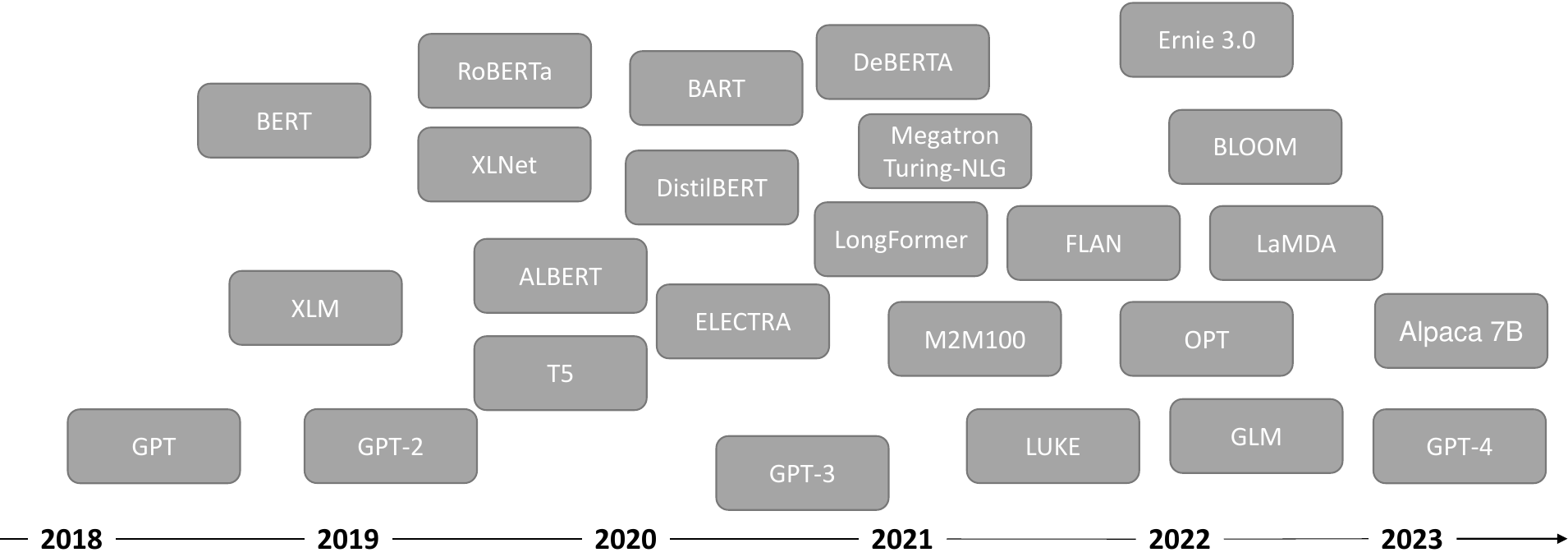}
\caption{Timeline for the development of transformer-based large language~models.\label{fig:history}}
\end{figure}

In this section, some of the leading transformer-based language models are presented. Figure~\ref{fig:history} displays the timeline of the evolution of LLMs since~2018.

\begin{figure}[h]
\centering
\includegraphics[width=\textwidth]{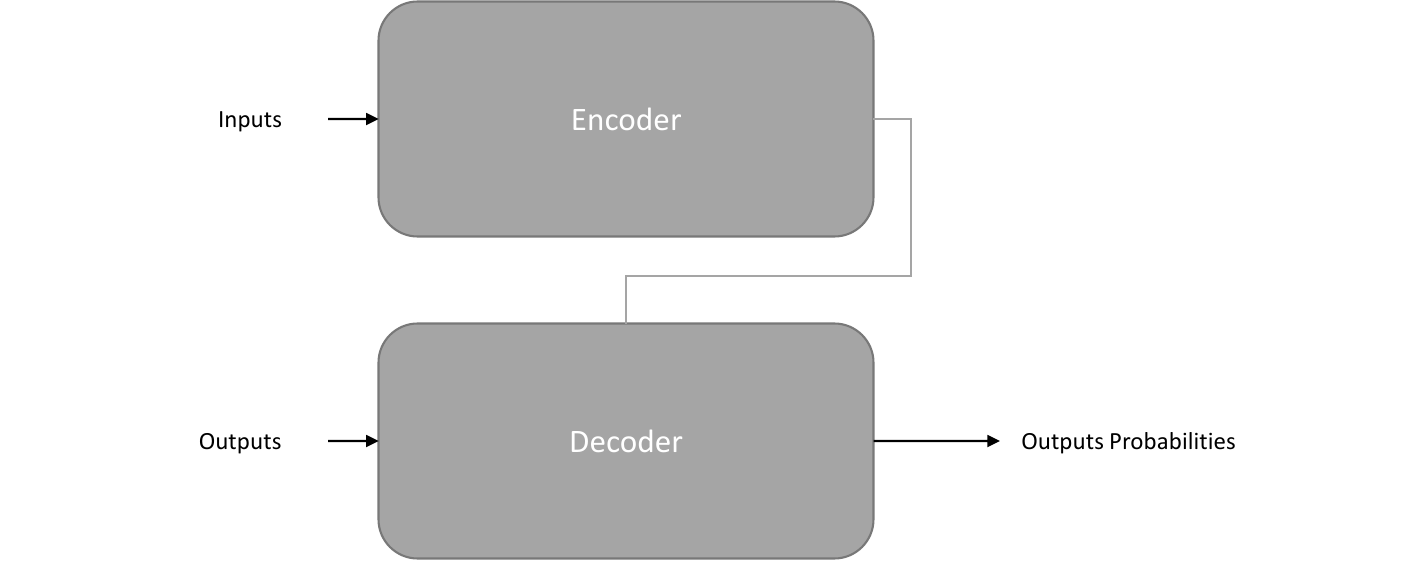}
% \end{minipage}
\caption{{\color{black}Encoder--decoder architecture}. The~model is primarily composed of two blocks: The encoder receives an input and builds a representation of its features, while the decoder uses the encoder’s representation along with other inputs to generate a target~sequence.}\label{fig:encoder_decoder}
\end{figure}

\begin{figure}[h]
\centering
\includegraphics[width=\textwidth]{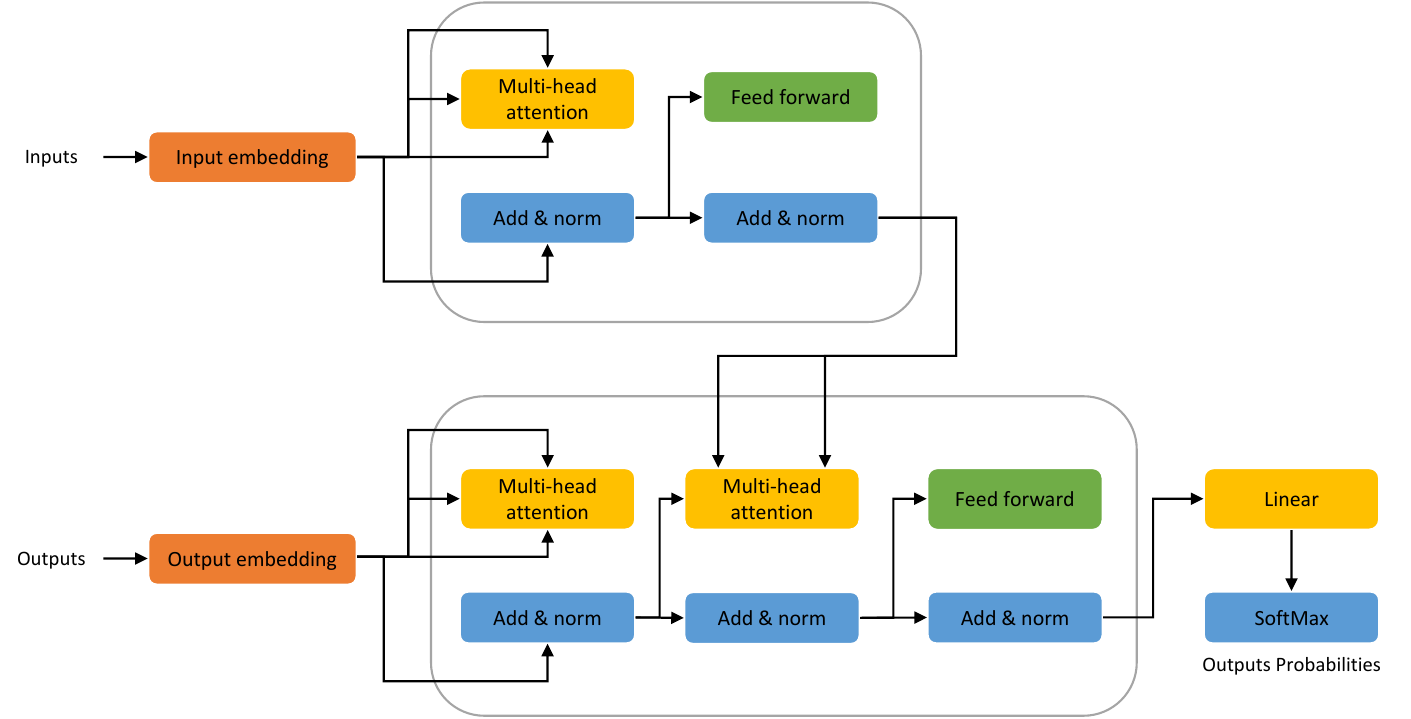}
\caption{{\color{black}Transformer architecture.} The transformer architecture retains a similar structure to that of the encoder--decoder architecture. The~encoder considers all words in a sentence, while the decoder works sequentially. Once the initial words are predicted, they are used to generate subsequent words. The~attention layers in the encoder consider all the words in a sentence, while the decoder works sequentially and can only focus on the words it has already~translated.}\label{fig:transformer}
\end{figure}

%\unskip  
Table~\ref{tab:llms} provides a summary of transformer-based language models used in AI-assisted programming. Transformer-based models are a type of neural network architecture used in NLP and other machine learning tasks. The~transformer maintains a similar architecture as the encoder--decoder architecture shown in Figure~\ref{fig:encoder_decoder}, but~the models use a self-attention mechanism to weigh the importance of different parts of the input sequence, allowing them to capture dependencies between all parts of the sequence, as~shown in Figure~\ref{fig:transformer}. They can be parallelized more easily than previous models, resulting in faster training and lower inference times. The~transformer model is one of the most well-known transformer-based models and has been used in various NLP tasks. Recently, large transformer-based models such as GPT-4~\cite{openai2023gpt4} and LLaMA~\cite{touvron2023llama} have achieved state-of-the-art performance in many benchmarks. The~transformer's ability to capture long-range dependencies is heavily reliant on dot-product attention with softmax normalization, leading to a quadratic space and time complexity in relation to sequence length, which can be a hindrance for longer inputs. This study focuses on transformer-based models for AI-assisted programming tasks. 

\begin{table}[!thb]
\caption{Summary of language models using transformers for AI-assisted~programming.\label{tab:llms}}
\begin{tabularx}{\textwidth}{CCC}
\toprule
\textbf{Model} & \textbf{Type}	& \textbf{AI-Assisted Programming Tasks}\\
\midrule
Encoder-only &  Understanding & Code Summarization, Code~Translation\\
Decoder-only & Generation & Code Generation, Code~Completion\\
Encoder--decoder   &  Generation and Understanding& Code Generation, Code Refinement, Defect Detection, Clone~Detection\\
\bottomrule
\end{tabularx}
\end{table}

\unskip
% \hfill

Encoder--decoder models~\cite{cho2014learning} refer to sequence-to-sequence models, utilizing both components of the transformer architecture~\cite{vaswani2017attention}. The~encoder's attention layers can access all words in the input sentence at each stage, while the decoder's attention layers can only access the words preceding a given word in the input. Sequence-to-sequence models such as BART~\cite{lewis2019bart}, T5 (Text-to-Text Transfer Transformer) \cite{raffel2020exploring}, and~TreeGen~\cite{sun2020treegen} are well-suited for tasks that involve generating new text based on an input, such as code generation, code refinement, defect detection, and~clone detection, for~AI-assisted programming~tasks.

Encoder-only models, also known as autoencoders, use only an encoder network to transform input data into a compressed representation. They are commonly used in unsupervised learning tasks such as dimensionality reduction and anomaly detection in NLP tasks. In~the past, code embedding approaches could be utilized to obtain the representation from the input data such as Neural Network Language Model~\cite{morin2005hierarchical}, Code2Vec~\cite{alon2019code2vec}, ELMo~\cite{peters2018deep}, TextRank~\cite{mihalcea2004textrank}, and~GGNN~\cite{allamanis2017learning}. For~AI-assisted programming tasks, they are used for understanding tasks to learn useful representations with the BERT~\cite{devlin2018bert} and RoBERTa~\cite{liu2019roberta} of data in an unsupervised manner, which can be used as features for downstream tasks such as code translation and code~summarization.

Decoder-only models, also known as autoregressive models, are a type of neural network architecture used in natural language processing tasks such as GPT-2~\cite{radford2019language}, GPT-3~\cite{brown2020language},
GPT-J~\cite{gpt-j}, Reformer~\cite{kitaev2020reformer}, and~GPT-Neo~\cite{black2022gpt}, which use the decoder to predict the next token output given all previous tokens. They rely solely on a decoder network to generate output text, predicting the probability distribution of the next token given the previously generated tokens. Although~they are simpler and more efficient than encoder--decoder models, they may not be as effective in tasks requiring a deeper understanding of the input--output sequence relationship. Nevertheless, they are still widely used in various natural language processing tasks for AI-assisted programming, such as code generation and code completion, and~have demonstrated impressive performance in several~benchmarks.

\subsection{Measurement of Language Models with~Entropy}\label{sec:elm}
Language models on software naturalness are trained on large code corpora and used to predict the next token in the code given its context. Mathematically, assuming a set of program tokens $\mathbb{T}$ and a set of program sequences $\mathbb{S}$, the~set of possible systems is $S \subset \mathbb{S}$. {A~language model is a probability distribution} $p(.)$ over systems $s \in S$:
\begin{align}
\label{eq:1}
    \forall s \in S [0 < p(s) <1] \land \sum_{s\in S} p(s) = 1.
\end{align}
An estimated language model known as a pre-trained language model~\cite{10.5555/555733} is created by computing a maximum-likelihood estimation (MLE) of the parameter of a suitably chosen parametric distribution $p(\cdot)$ given a corpus $C$ of programs $C \subseteq S$. This process is described in Section~\ref{sec:token}. The tokenization of the code is defined by the programming language to estimate the probability distribution of code tokens given the preceding context. It uses this information to make predictions or decisions in the software engineering tasks. The~models are trained to predict the probability distribution of words in a sequence, based on the previous words in that sequence~\cite{bengio2000neural}. The~language model is typically constructed using $N$-gram models, which have a long history in statistical language modeling and are widely used for estimating the probability distribution of words or characters in a text sequence~\cite{katz1987estimation,brown1992class}. This was the standard method before the development of word vectors and distributed representations of language using Recurrent Neural Networks (RNN) \cite{mikolov2013efficient}. Given a system $s$ with a sequence of tokens $\{W_1,W_2,\dots W_n\}$, $N$-gram models can estimate the likelihood of tokens following other tokens. As~a result, the~model can estimate the probability of $s$ by multiplying a series of conditional probabilities:
\begin{align}
\label{eq:2}
    p(s) = p(W_1)p(W_2|a_1)p(W_3|W_1 W_2)\dots p(W_n|W_1\dots W_{n-1}).
\end{align}
An $N$-gram model captures the co-occurrence patterns of words or characters in the text. Mathematically, an~$N$-gram model can be represented as a set of $N$-grams, each represented as a tuple of $n$ items and their associated probabilities. The~probability of an $N$-gram can be estimated by the MLE based on the frequency of occurrence of the $N$-gram in a given training corpus. This also assumes a Markov property, i.e.,~token occurrences are influenced only by a limited prefix length of $n$. Thus, for~example, \linebreak in~a $3$-gram $(n=3)$ model:
\begin{align}
\label{eq:3}
    p(W_i|W_1 \dots W_{i-1}) \cong p(W_i | W_{i-2} W_{i-1}).
\end{align}
The probability of a word $W_i$ given its preceding word $W_{i-1}$ can be estimated:
\begin{align}
\label{eq:4}
p(W_i | W_{i-1}) = count(W_{i-1}, W_i) / count(W_{i-1}),
\end{align}
where $count(W_{i-1}, W_i)$ is the number of times the $3$-gram $(W_{i-1}, W_i)$ appears in the training corpus, and~$count(W_{i-1})$ is the number of times the word $W_{i-1}$ appears in the training corpus. The~models have achieved great success in recent years and have been a driving force behind recent advancements in NLP. The~performance of the technique depends on the quality of the language model and the ability of the model to accurately reflect the patterns and structures of the target data. Therefore, much research effort has been devoted to improving the quality of language models for these tasks, including developing better training algorithms, larger training corpora, and~better evaluation~metrics.

A representative corpus of repetitive and highly predictable programs is utilized to capture regularities within the corpus in order to evaluate the naturalness of software language models. By estimating the language model from this representative corpus, it can predict the contents of new programs with high confidence, thereby minimizing the surprise associated with the new program. In~NLP, this idea is often measured using perplexity or cross-entropy (log-transformed version). Given a program $p = \{w_1,w_2,\dots,w_n\}$, of~length~$n$, and~a language model $\Theta$, it assumes that the probability of the programs estimated by the model is $p_{\Theta}$, and, thus, the~cross-entropy $H_{\Theta}(p)$ can be measured:
\begin{align}
\label{eq:5}
    H_{\Theta}(p) = - \frac{1}{n} \log p_\Theta (w_1,w_2,\dots, w_n)
\end{align}
and a formulation can be derived from Equation~(\ref{eq:2}):
\begin{align}
\label{eq:6}
    H_{\Theta}(p) = - \frac{1}{n} \sum^n_{i=1} \log p_\Theta (w_i|w_1,w_2,\dots,w_{i-1}).
\end{align}
The entropy rate of a language model is utilized to assess the naturalness of the generated text~\cite{shannon1951prediction}. It can be computed by taking the negative logarithm of the probability of each generated token. An~effective model should have low entropy for the majority of programs, assigning higher probabilities (i.e., values closer to 1) to most words in the program, thereby resulting in lower absolute log values. In~practice, this involves using techniques such as maximum likelihood estimation or neural networks to estimate the parameters. The~final model can then be used to make predictions by calculating the probability of a given sequence of words. Estimating entropy from empirical data has been an interesting area in information theory for AI-assisted programming~\cite{mozannar2022reading}. For~example, a~method for estimating entropy with a confidence interval was proposed in~\cite{ho2010interplay}. Another method for estimating the entropy and redundancy of a language was provided in~\cite{shannon1951prediction}. A~model weighting principle based on the minimum description length principle was applied in~\cite{kennel2005estimating} to develop a direct estimator of the entropy rate. The~estimator can be used to estimate a Bayesian confidence interval for the entropy rate using Monte Carlo techniques. Techniques for estimating the entropy rate have been reviewed in~\cite{feutrill2021review}. Analytical results of estimators for entropy and mutual information can be found in~\cite{paninski2003estimation}.
\section{AI-Assisted Programming~Tasks}\label{sec:task}
There are two main categories of AI-assisted programming tasks related to software naturalness: generation and understanding. The~former includes code generation, code completion, code translation, code refinement, and~code summarization. The~latter is concerned with understanding code and includes defect detection and clone detection. Researchers have made significant efforts to enhance the quality of language models for these tasks by improving pre-training schemes, increasing the size of training corpora, developing better fine-tuning datasets, and~using improved evaluation metrics. The~frameworks and tools developed for these specific tasks are discussed in this section, and~a summary of all the frameworks reviewed is presented in Table~\ref{tab:tasks}.

\subsection{Code~Generation} \label{sec:cg}
Program synthesis, also known as source code generation, is the process of automatically generating source code from a programming language based on user-specified constraints~\cite{waldinger1969prow,manna1971toward}. This study focuses on text-to-code generation for code generation, while code-to-code generation is referred to as code translation, which is discussed \mbox{in Section~\ref{sec:ct}.} The~history of code generation dates back to the use of theorem provers to construct a proof of user-provided specifications and extract corresponding logical programs~\cite{manna1975knowledge, green1981application}. With~the increasing popularity of deep learning methods, neural methods, including Long Short--Term Memory (LSTM)~\cite{dong2016language} and Recursive--Reverse--Recursive Neural Network~\cite{parisottoneuro}, have been adopted to generate output programs with specific inductive biases given sufficient program samples. More recently, transformer-based LLMs such as GPT-3~\cite{brown2020language} and T5~\cite{raffel2020exploring} have shown impressive performance in code generation tasks by leveraging contextual representations learned from large amounts of code, as~well as public code sources and natural language data, to~improve program synthesis. These approaches incorporate systematic pre-training and fine-tuning tasks to develop a deep understanding of code structure and meaning, making them well-suited for software development tasks. To~evaluate the models for code generation tasks, different metrics are available such as $pass@k$~\cite{chen2021evaluating}, which measures the percentage of problems solved using $k$ generated programs per problem, BLEU-4~\cite{lin2004orange}, and~exact match accuracy on program synthesis benchmarks such as APPS~\cite{hendrycksapps2021}, MBPP~\cite{austin2021program}, and~CodeBLEU~\cite{raffel2020exploring}, which consider both syntactic and semantic matches based on code structure in addition to $N$-gram~matches.

\subsection{Code~Completion}\label{sec:cc}
Code completion, also known as autocompletion, is a software development feature that suggests possible code completions as a programmer types~\cite{dong2022snr}. Its goal is to save time and reduce errors by providing suggestions for method names, variable names, and~even entire code snippets~\cite{amazon}. Previous research on code completion started with statistical language models~\cite{robbes2008program,bruch2009learning}. Later, LSTM-based deep learning approaches were applied to the task, aiming to learn the semantic information of source code without considering its syntactic structure~\cite{svyatkovskiy2019pythia}. To~address the limitations of LSTM-based language models, transformer architecture was introduced for code completion. Normally, the~language models for code completion are trained using a causal language model that predicts the unknown token after a sequence of known tokens. Recent work on code completion using LLMs~\cite{takerngsaksiri2022syntax, chen2021evaluating} has shown impressive performance on benchmarks, such as CodeXGLUE~\cite{lu1codexglue}, compared to existing statistical language models and deep learning~approaches.

\subsection{Code~Translation}\label{sec:ct}
Code translation is the process of converting code from one programming language to another, with~the goal of migrating legacy software. While theoretically possible, building a code translator is challenging due to differences in syntax and platform APIs between programming languages. Most current translation tools are rule-based, requiring handcrafted rewrite rules applied to an abstract syntax tree (AST) derived from the input source code. However, creating such tools demands significant expertise in both the source and target languages. Recent studies have explored using statistical machine translation~\cite{koehn2007open, artetxeetal2018unsupervised} as well as deep learning approaches~\cite{allamanis2014naturalize, acharya2007mining} for programming language translation. Quality evaluation for generated functions often uses the BLEU score, while the exact match is used to compare generated output with reference ground~truth.

\subsection{Code~Refinement}\label{sec:cr}
Code refinement, which can be referred to as automated program repair (APR), is the process of automatically fixing bugs or vulnerabilities by converting a buggy function into a correct one. Deep learning models have a strong learning capability that enables them to learn various patterns for transforming buggy programs into patched ones from large code corpora. Many studies~\cite{jiang2021cure,zhu2021syntax} have demonstrated the superior performance of deep learning-based techniques over traditional template-based~\cite{jiang2018shaping,liu2019tbar}, heuristic-based~\cite{yuan2018arja,wen2018context,saha2017elixir}, and~constraint-based~\cite{xiong2017precise,xuan2016nopol} APR techniques. LLM is used to generate plausible patches or modifications to a given incorrect code. The~model can be trained on a large corpus of correct code to learn the patterns and structures of correct code. When LLMs are given a faulty code, the~model can then generate suggestions for how to correct it as one of the downstream tasks. The LLMs for code refinement can be evaluated by CodeXGLUE~\cite{lu1codexglue} or HumanEval~\cite{chen2021evaluating} as the abstracted codes or the classical APR benchmarks such as Defects4J~\cite{just2014defects4j} and QuixBugs~\cite{lin2017quixbugs} as real-world codes, but~the understanding and generation of concrete variable and function names is still mandatory and challenging~\cite{jiang2023impact}.

\subsection{Code~Summarization}
Code summarization is a technique used to generate English descriptions of code snippets at the function level, which can then be used to generate documentation. Typically, this involves taking the source code as input and producing a natural language summary as output. In~AI-assisted programming tools, code summarization can be used to analyze code and identify optimization opportunities, such as using a binary Euclid algorithm instead of a traditional modular arithmetic-based algorithm, which can significantly improve software performance. In~recent years, there has been promising research into the automatic generation of natural language descriptions of programs, with~studies such as~\cite{sridhara2010towards,moreno2013automatic,sridhara2011generating} making notable progress in this area. The~rise of deep learning, coupled with the abundance of data from open-source repositories, has made automatic code summarization an area of interest for researchers. Many of the neural approaches~\cite{ahmad2020transformer, iyer2016summarizing} use a sequence-to-sequence approach to generate source code summaries, with~some models converting the source code into various types of representations, such as token-based~\cite{allamanis2016convolutional, chen2018neural}, tree-based~\cite{mou2016convolutional, liang2018automatic}, and~graph-based~\cite{tufano2018deep, ou2016asymmetric}, before~passing it through language~models.

\subsection{Defect~Detection}\label{sec:dd}
As software systems increase in complexity, it becomes more challenging to identify errors. Defect detection aims to enhance software reliability by predicting whether a piece of code is susceptible to bugs or not, by~detecting previously unknown errors. Rule-based approaches have been defined in existing defect detection frameworks by inferring likely programming rules from various sources such as code, version histories, \mbox{and~comments~\cite{acharya2007mining,livshits2005dynamine,wasylkowski2007detecting}.} Statistical language models based on $N$-gram language models have also been widely used in this area~\cite{charniak1996statistical,nessa2008software,wang2016bugram}. More recently, many deep learning-based solutions~\cite{lin2018cross, li2018vuldeepecker, russell2018automated,le2019maximal,chen2019sequencer,gupta2017deepfix, liu2019tbar} have been proposed to bridge the gap by suggesting different feature sets from which the detection framework can learn, attempting to imitate how a practitioner looks for vulnerabilities. However, LLMs, such as CodeBERT~\cite{feng2020codebert}, have recently emerged as a promising technique in this field due to their ability to understand code structure. These models can be trained on a large corpus of error-free code and used to identify patterns and structures in source code that deviate from those learned from the error-free code as a binary classification task~\cite{buratti2020exploring, li2022automating}. To~evaluate the model predictions, accuracy, precision, recall, and~F1 scores can be~used.

\subsection{Clone~Detection}\label{sec:cd}
Clone detection involves identifying identical or similar code fragments, known as clones, within~or across software systems. The~goal of clone detection is to measure the similarity between two code snippets and determine if they have the same functionality. Clones can be classified into four types~\cite{bellon2007comparison,roy2007survey}, with~types 1--3 being syntactic clones that differ in minor ways, while type 4 clones, known as semantic clones, are difficult to detect since they have different syntax but the same semantics and, thus, require manual validation. With~the increasing amount of source code, large-scale and automatic clone detection has become essential. Several tools have been developed to perform clone detection~\cite{kontogiannis1996pattern,ducasse1999language, baxter1998clone,chen2014achieving,sajnani2016sourcerercc, yu2019neural}, using techniques such as comparison of the AST, tokens, or~source code text. Notable clone detection datasets include BigCloneBench~\cite{svajlenko2021bigclonebench}, which contains Java code~snippets.

\begin{table}[!thb]
    \caption{Summary of language models for AI-assisted programming tasks.}\label{tab:tasks}
 \begin{tabularx}{\textwidth}{p{2.5cm}<{\raggedright}p{1cm}<{\raggedright}p{3cm}<{\raggedright}p{1.9cm}<{\raggedright}p{2cm}<{\raggedright}p{2cm}<{\raggedright}p{1.5cm}<{\raggedright}}
    \toprule
	{\textbf{Framework}} & {\textbf{Year}} & {\textbf{Task(s)}} & {\textbf{Baseline(s)}}& {\textbf{Supported Language(s)}}& {\textbf{Open Sourced}}\\
	\midrule
    Refactory~\cite{yang2019refactory} & 2019 & Defect Detection& BLEU& Java & \xmark \\
    CuBERT~\cite{kanade2020learning} & 2020 & Code Refinement, Defect Detection &BERT & Python & \cmark\\
    CugLM~\cite{liu2020multi} &2020 & Code Completion & BERT & Java, TypeScript & \cmark\\
    Intellicode~\cite{svyatkovskiy2020intellicode}& 2020 & Code Generation, Code Completion& GPT-2&  Python, C\#, JavaScript, and~TypeScrip &\xmark\\
    Great~\cite{hellendoornglobal} &2020 &Defect Detection & Vanilla Transformers & Python &\cmark\\
    TreeGEN~\cite{sun2020treegen} &2020&  Code Generation & Vanilla Transformers &  Python &  \cmark\\
    C-BERT~\cite{buratti2020exploring} & 2020 & Defect Detection & BERT&  C & \xmark\\
    TransCoder~\cite{roziere2020unsupervised} & 2020 & Code Translation & Vanilla Transformers & C++, Java, and~Python & \xmark\\
    GraphCode- BERT~\cite{guo2020graphcodebert} &2020 & Code Summarization, Code Refinement & BERT & Java & \xmark\\ 
    Codex~\cite{chen2021evaluating} &2021& Code Generation, Code Completion, Code Summarization, Benchmark& GPT-3 & JavaScript, Go, Perl, and~6 more & \xmark\\
    Copilot~\cite{friedman2021introducing} & 2021 & Code Generation, Code Completion & Codex & Java, PHP, Python, and~5 more &\xmark\\
    BUGLAB~\cite{allamanis2021self} & 2021 & Code Refinement, Defect Detection &GREAT& Python & \cmark\\
    TBCC~\cite{hua2022transformer} & 2021 & Clone Detection &Vanilla Transformers & C, Java & \cmark\\
    \bottomrule
	\end{tabularx}
\end{table}
\begin{table}[!thb]\ContinuedFloat
    \caption{Cont.}
 \begin{tabularx}{\textwidth}{p{2.5cm}<{\raggedright}p{1cm}<{\raggedright}p{3cm}<{\raggedright}p{1.9cm}<{\raggedright}p{2cm}<{\raggedright}p{2cm}<{\raggedright}p{1.5cm}<{\raggedright}}
    \toprule
	{\textbf{Framework}} & {\textbf{Year}} & {\textbf{Task(s)}} & {\textbf{Baseline(s)}}& {\textbf{Supported Language(s)}}& {\textbf{Open Sourced}}\\
	\midrule
    CodeT5~\cite{wang2021codet5} &2021& Code Summarization, Code Generation, Code Translation, Code Refinement, Defect Detection, Clone Detection& T5 & Python, Java & \cmark\\
    Tfix~\cite{berabi2021tfix} & 2021 & Code Refinement, Defect Detection & T5 & JavaScript & \cmark\\
    CodeRL~\cite{lecoderl} &2021& Code Summarization, Code Generation, Code Translation, Code Refinement, Defect Detection, Clone Detection & T5 & Java & \cmark\\
    TreeBERT~\cite{jiang2021treebert} & 2021 &Code Summarization &Vanilla Transformers &Python, Java&\cmark\\
    
    APPS~\cite{hendrycksapps2021} & 2021 & Benchmark & N/A & Python & \cmark\\
    CodeXGLUE~\cite{lu1codexglue} & 2021 & Benchmark & N/A & Python & \cmark\\
    CoTexT~\cite{phan2021cotext} & 2021 & Code Summarization, Code Generation, Code Refinement, Defect detection & T5& Python, Java, Javascript, PHP, Ruby, Go & \cmark\\
    SynCoBERT~\cite{wang2021syncobert} &2021 & Code Translation, Defect Detection, Clone Detection & BERT & Ruby, Javascript, Go, Python, Java, PHP & \xmark\\
    TravTrans~\cite{kim2021code} & 2021 &Code Completion & Vanilla Transformers&Python& \xmark\\
    CCAG~\cite{wang2021code} & 2021 & Code Completion & Vanilla Transformers & JavaScript, Python & \xmark\\
    DeepDebug~\cite{drain2021deepdebug}&2021 & Defect Detection & Reformer & Java & \cmark \\
    Recoder~\cite{zhu2021syntax}&2021& Defect Detection & TreeGen & Java & \cmark \\
    PLBART~\cite{ahmad2021unified} & 2021 & Code Summarization, Code Generation, Code Translation, Code Refinement, Clone Detection, Detect Detection & BART &Java, Python & \xmark\\
    CODEGEN~\cite{Nijkamp2022CG} & 2022& Code Generation & GPT-NEO \& GPT-J &Python &\cmark \\
    GPT-2 for APR~\cite{lajko2022towards} & 2022 & Code Refinement & GPT-2 & JavaScript & \cmark\\
    CERT~\cite{CERT} & 2022 & Code Generation & CODEGEN & Python & \cmark\\
    \bottomrule
	\end{tabularx}
\end{table}
\begin{table}[!thb]\ContinuedFloat
    \caption{Cont.}
 \begin{tabularx}{\textwidth}{p{2.5cm}<{\raggedright}p{1cm}<{\raggedright}p{3cm}<{\raggedright}p{1.9cm}<{\raggedright}p{2cm}<{\raggedright}p{2cm}<{\raggedright}p{1.5cm}<{\raggedright}}
    \toprule
	{\textbf{Framework}} & {\textbf{Year}} & {\textbf{Task(s)}} & {\textbf{Baseline(s)}}& {\textbf{Supported Language(s)}}& {\textbf{Open Sourced}}\\
	\midrule
    PyCoder~\cite{takerngsaksiri2022syntax} & 2022 & Code Generation & GPT-2 &Python&\cmark\\
    AlphaCode~\cite{li2022competition}&2022 & Code Generation & GPT & Java & \xmark\\
    InCoder~\cite{fried2022incoder}&2022 & Code Generation, Code Completion, Code Summarization& GPT-3 & Java, JavaScript, Python & \cmark\\
    RewardRepair~\cite{ye2022neural}&2022 & Code Refinement, Defect Detection & T5 & Java & \cmark\\
    CodeParrot~\cite{tunstall2022natural}& 2022 & Code Generation  & GPT-2 & Python &\cmark\\
    AlphaRepair~\cite{xia2022less} & 2022 & Code Refinement, Defect Detection & CodeBERT  & Java & \cmark\\
    CodeReviewer~\cite{li2022automating} & 2022 &Code Summarization, Code Refinement, Defect Detection & CodeT5 & Java & \cmark\\
    TransRepair~\cite{li2022transrepair}&2022 & Code Refinement, Defect Detection & BLEU & Java & \xmark \\
    NatGen~\cite{chakraborty2022natgen} & 2022 & Code Generation, Code Translation, Code Refinement & CodeT5 & Java, Python, Go, JavaScript, Ruby, PHP & \cmark\\
    DualSC~\cite{yang2022dualsc} &2022 & Code Generation, Code Summarization & T5 &Shellcode & \cmark\\
    VulRepair~\cite{fu2022vulrepair} & 2022 & Code Refinement, Defect Detection & T5 & C, C++ & \cmark\\
    CoditT5~\cite{ZhangETAL22CoditT5} &2022 & Code Summarization, Defect Detection &CodeT5 & Java, Python, Ruby, PHP, Go, JavaScript & \cmark \\
    C4~\cite{tao2022c4} & 2022 & Clone Detection & CodeBERT & C++, C\#, Java, Python & \cmark\\
    SPT-Code~\cite{niu2022spt} & 2022 &Code Summarization, Code Completion, Code Refinement,  Code Translation & CodeBERT \& GraphCodeBERT & Python, Java, JavaScript, PHP, Go & \cmark\\
    ExploitGen~\cite{yang2023exploitgen} & 2023 & Code Generation & CodeBERT & Python, Assembly & \cmark\\
    Santacoder~\cite{allal2023santacoder}& 2023 &  Code Summarization, Code Generation&GPT-2 & Python, Java, and~Javascript & \cmark\\
xCodeEval~\cite{khan2023xcodeeval} &2023& Benchmark& N/A & Python, Java, C++, PHP, and~8 more &\cmark\\
StarCoder~\cite{li2023starcoder} & 2023 &  Code Generation, Code Completion, Code Summarization & BERT \& SantaCoder &HTML, Python, Java, and~83 more&\cmark\\
    \bottomrule
	\end{tabularx}
\end{table}

\section{Challenges and~Opportunities}\label{sec:chop}
\unskip
\subsection{Computational~Expense}
Training an LLM with millions of parameters can be computationally expensive. This is because training involves processing vast amounts of data in codes and optimizing the model's parameters to generate accurate predictions~\cite{zhang2020accelerating}. Overall, computational expense can be due to lack of training data and computing resources such as memory, GPU, or~even electricity. At~the same time, the~quality of the training data used to train a language model is also crucial, as~poor quality data or bias in the data can lead to incorrect predictions. LLMs require massive computational resources to train, fine-tune, and~run, which can be a hindrance for organizations with limited hardware resources~\cite{han2021pre}. 

To reduce the computational expense of training LLMs, researchers and developers can employ various techniques, such as training on subsets of the data~\cite{lin2009select,liang2022metashift}, optimizing the hyperparameters~\cite{yin2021autotinybert}, and~leveraging transfer learning to reuse the knowledge learned from previous tasks. These techniques can help to speed up the training process and reduce {\color{black}the amount of required computing resources.} Instead of training the LLMs continuously, some works focus on using prompt-learning~\cite{openai_2023,serban2017deep} and human feedback~\cite{christiano2017deep,ling2018human,ziegler2019fine,stiennon2020learning,ouyang2022training} to improve performance of the LLMs. In~prompt-based learning, the~prompt serves as a guide or prompt to the language model, providing it with relevant context and guidance to generate an output that is appropriate for a particular task. The~prompt can be a simple sentence or a full paragraph, depending on the complexity of the task and the amount of information needed to guide the LLMs. One of the main advantages of prompt-based learning is its flexibility and ease of use. It allows users to quickly fine-tune pre-trained language models for specific tasks without requiring a large amount of task-specific data. Additionally, prompt-based learning can be used in a semi-supervised or unsupervised manner, where the prompt provides a small amount of supervision to the language model, {\color{black}further reducing the necessary amount of task-specific data.}

\subsection{Quality~Measurement}
Leveraging LLMs in AI-assisted programming tasks has enormous potential to improve software development efficiency and reduce the time and effort required to write code manually. However, several challenges need to be addressed to ensure the performance and effectiveness of LLMs. One of the primary concerns is the quality of the generated code or documentation~\cite{chen2021evaluating}, which can be impacted by the accuracy and robustness of the LLMs. While automated code generation can save time, it can also lead to poor-quality code that is difficult to maintain and may contain bugs or security vulnerabilities~\cite{hendler2023understanding}. Therefore, it is critical to ensure that the generated code meets the desired specifications and adheres to coding standards and best practices~\cite{chen2022codet}. Another significant challenge is integrating the generated code into existing software systems seamlessly~\cite{white2023assessment}, ensuring that it can be maintained and updated easily over~time.

To address these challenges and improve the reliability and quality of LLMs in \mbox{AI-assisted} programming tasks, researchers and developers are exploring various approaches and techniques. These include incorporating advanced {\color{black}machine learning and optimization} algorithms~\cite{howard2018universal, wei2021finetuned} and developing new tools and frameworks for integrating generated code into existing software systems. Some researchers have attempted to use Variational Autoencoders~\cite{kingma2013auto} or Generative Adversarial Networks~\cite{goodfellow2020generative} to generate synthetic data that can be used for training LLMs, but~they must ensure that the performance of these generative models is robust and reliable to ensure the quality of the synthetic data. Meanwhile, {\color{black}it is possible to adopt active learning} \cite{settles2009active} to improve the performance of LLMs while requiring fewer labeled training instances. This approach works by allowing the model to choose the data {\color{black}from which it learns~\cite{cohn1996active}, which} enables it to compute the statistically optimal way to select training data while avoiding poor-quality data, such as buggy codes, that can negatively impact model performance. One of the significant benefits of incorporating active learning into the training process is that it can help reduce the time and effort required to label large amounts of data manually, making it a cost-effective solution for many applications~\cite{settles2008active}. By~selecting the most informative data points for labeling, active learning can improve the accuracy and robustness of machine learning models, even when working with limited labeled data. The~integration of active learning with LLMs remains an open question in this field of study. While active learning has shown promise in improving the performance of machine learning models, including LLMs, the~application of this technique to LLMs has not yet been fully~explored.

\subsection{Software~Security}
Software security is a critical concern in the development of the use of LLMs~\cite{he2023large}. While LLMs have shown significant promise in a wide range of code-related tasks, they also introduce unique security challenges that must be addressed to ensure safety and security. One of the primary security concerns when using LLMs is the potential for these models to introduce vulnerabilities into the code~\cite{pearce2022asleep}. For~example, poorly designed LLMs may generate code that is prone to buffer overflow or SQL injection attacks. Another critical concern is the possibility of LLMs being manipulated or exploited to generate malicious code that can be used for cyberattacks. For~instance, an~attacker may use a poisoned dataset to manipulate an LLM, resulting in the generation of malicious code that can be used to exploit vulnerabilities in the software system. Also, users without programming knowledge can generate programs with a Trojan horse phishing~attack.

When using LLMs for AI-assisted programming tasks, it is essential to address software security to ensure that the generated codes or documents are secure and free from vulnerabilities, as~well as to ensure the integrity of the training data used to train the LLMs. Code validation and testing involve thorough validation and testing of the generated code before integrating it with real-world systems to identify and fix any security issues. Data sanitization and validation ensure that the training data are free from malicious code or sources of~bias.

\subsection{Software~Piracy} 
Software piracy refers to the unauthorized copying, distribution, or~use of copyrighted software without the permission of the software's owner~\cite{peace2003software, reavis1991software, limayem2004factors}. This can take many forms, including making copies of software for personal or commercial use, distributing software through unauthorized channels, or~using software beyond the terms of the licensing agreement. As~the field of natural language generation and statistical machine learning for Big Code and AI-assisted programming continues to grow, concerns over software piracy have arisen. The~use of open source code repositories for training AI~models has led to lawsuits, with~companies such as Microsoft and OpenAI accused of software piracy. The~issue at hand is whether the use of open source code for training LLMs violates copyright laws. While the legal implications of this issue are still being debated, it is important to consider the ethical implications as well. The~use of copyrighted code without permission raises questions about fairness and equity in the development of AI-assisted programming tools~\cite{de2005copyright, kelty2004culture}. Also, the~use of user data to train these models raises concerns over privacy and data protection. As~the field continues to evolve, it will be important for researchers and developers to consider these issues and work towards finding solutions that balance the benefits of AI-assisted programming with the need for ethical and legal compliance. This may include clarifying rules around secondary uses of copyrighted code, as~well as developing more transparent and opt-in data policies for training AI~models. 

To address software piracy, one approach is to ensure that the training data used for the development of these models are legally obtained and do not violate any copyrights or intellectual property rights according to the U.S. Copyright Office~\cite{libraryofcongress_2023}. Organizations can also establish clear policies and guidelines for the ethical and legal use of these technologies. For~instance, developers can be required to obtain permission or licenses before using proprietary code or software in their work. Machine learning algorithms can also be trained to identify and prevent the unauthorized distribution of copyrighted material and pirated code or~software.

\subsection{Integration with Existing~Tools}
\begin{figure}[!thb]
\centering
\includegraphics[width=\linewidth]{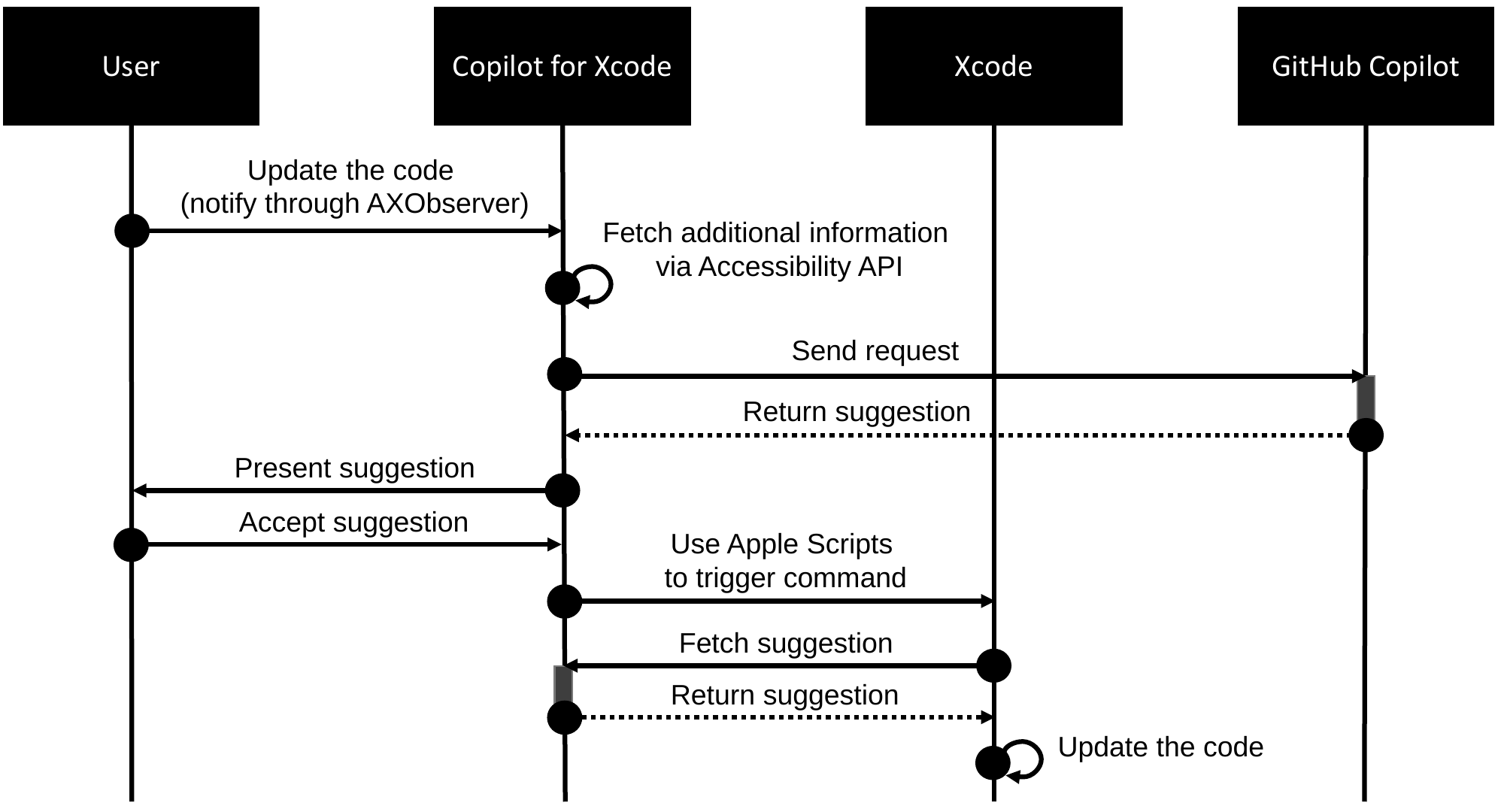}
\caption{A sequence diagram of Copilot for Xcode to produce real-time suggestions with GitHub Copilot. When a user attempts to update their code, the~Copilot for Xcode first receives a notification and sends a request to the GitHub Copilot API. Once the suggestions from GitHub Copilot are returned, the~user can choose to adopt the suggestions and apply the changes directly to~Xcode.}\label{fig:copilot_for_xcode_fetch_suggestions}
\end{figure}

The opportunity to integrate tools and LLMs enhances and streamlines the software development process. By~incorporating LLMs into integrated tools as cloud virtual service providers~\cite{zheng2015bid,zheng2016viability}, developers can leverage the power of NLP to automate repetitive tasks, improve code quality and readability, and~increase efficiency in software development. This integration can enable developers to experiment prompt engineering with public LLMs under data compliance, data security, data governance and best practices directly from their own development environment. Copilot for Xcode~\cite{guo} serves as a real-world example of an application integrated with LLMs, allowing Apple developers to utilize GitHub Copilot~\cite{friedman2021introducing} for code suggestions and ChatGPT~\cite{openai_2023} for code explanation and mutation using natural language. The~connection between Xcode and Copilot is achieved by establishing communication between the Xcode source editor extension and the Copilot server, presenting suggestions in a user interface not handled by Xcode. To~obtain additional information beyond the source code and file type provided by Xcode, the~app utilizes the Accessibility API, which represents objects in a user interface and exposes information about each object within the application. Furthermore, for~in-place code editing, the~app employs the use of Apple Scripts, a~scripting language in macOS for task automation, to~programmatically execute extension commands and emulate menu bar interactions. The~details to integrate the Copilot with Xcode are illustrated in Figure~\ref{fig:copilot_for_xcode_fetch_suggestions}.

\begin{figure}[!thb]
	\centering
	\begin{subfigure}{\textwidth}
		\centering
		\includegraphics[width=\linewidth]{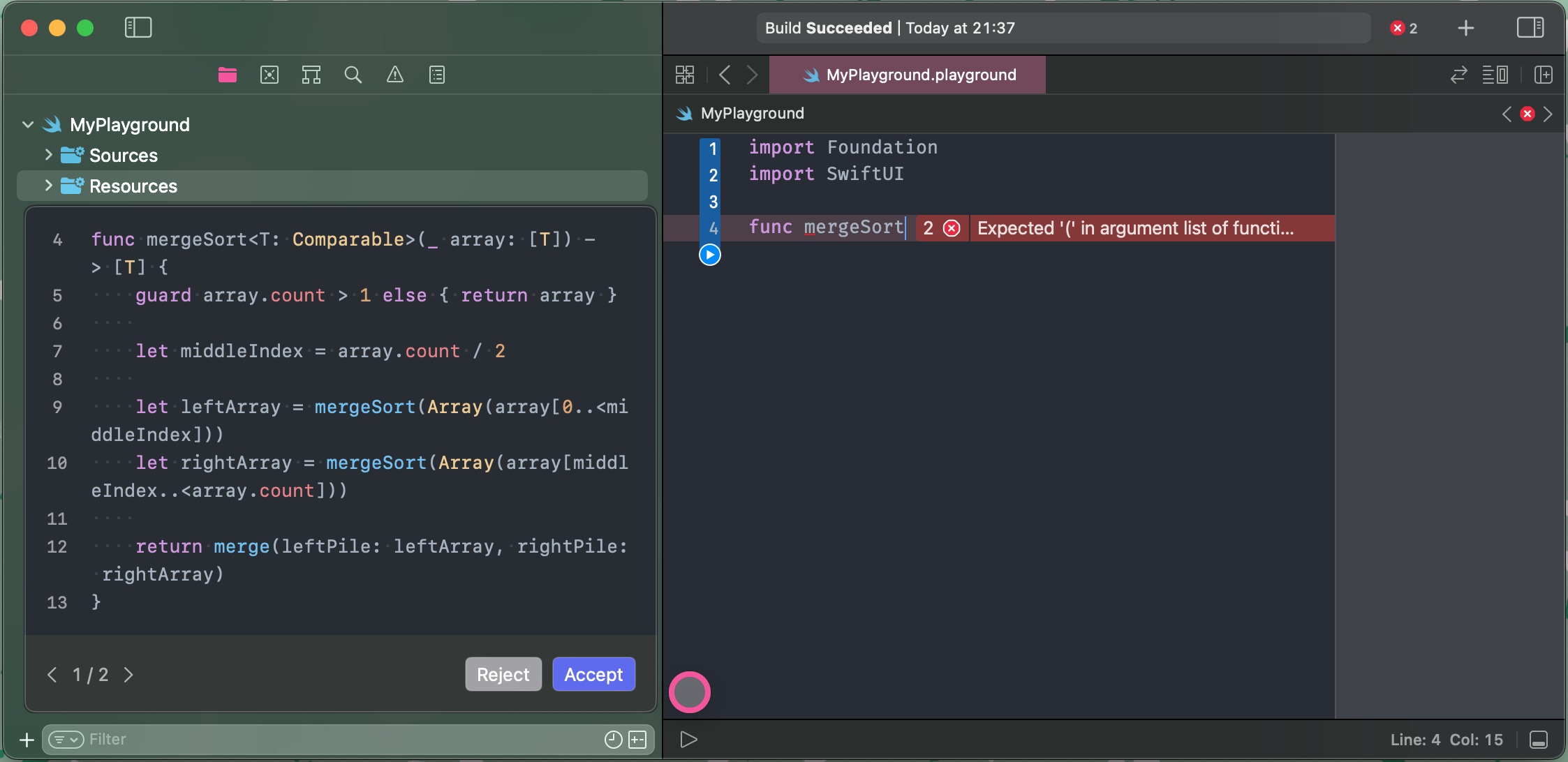}
		\captionsetup[subfigure]{justification=centering}
		\caption{Copilot for Xcode displaying suggestions from GitHub~Copilot.}
		\label{fig:copilot_for_xcode_screenshot_1}
        \hspace{2cm}
	\end{subfigure}
    
	\begin{subfigure}{\textwidth}
		\centering
		\includegraphics[width=\linewidth]{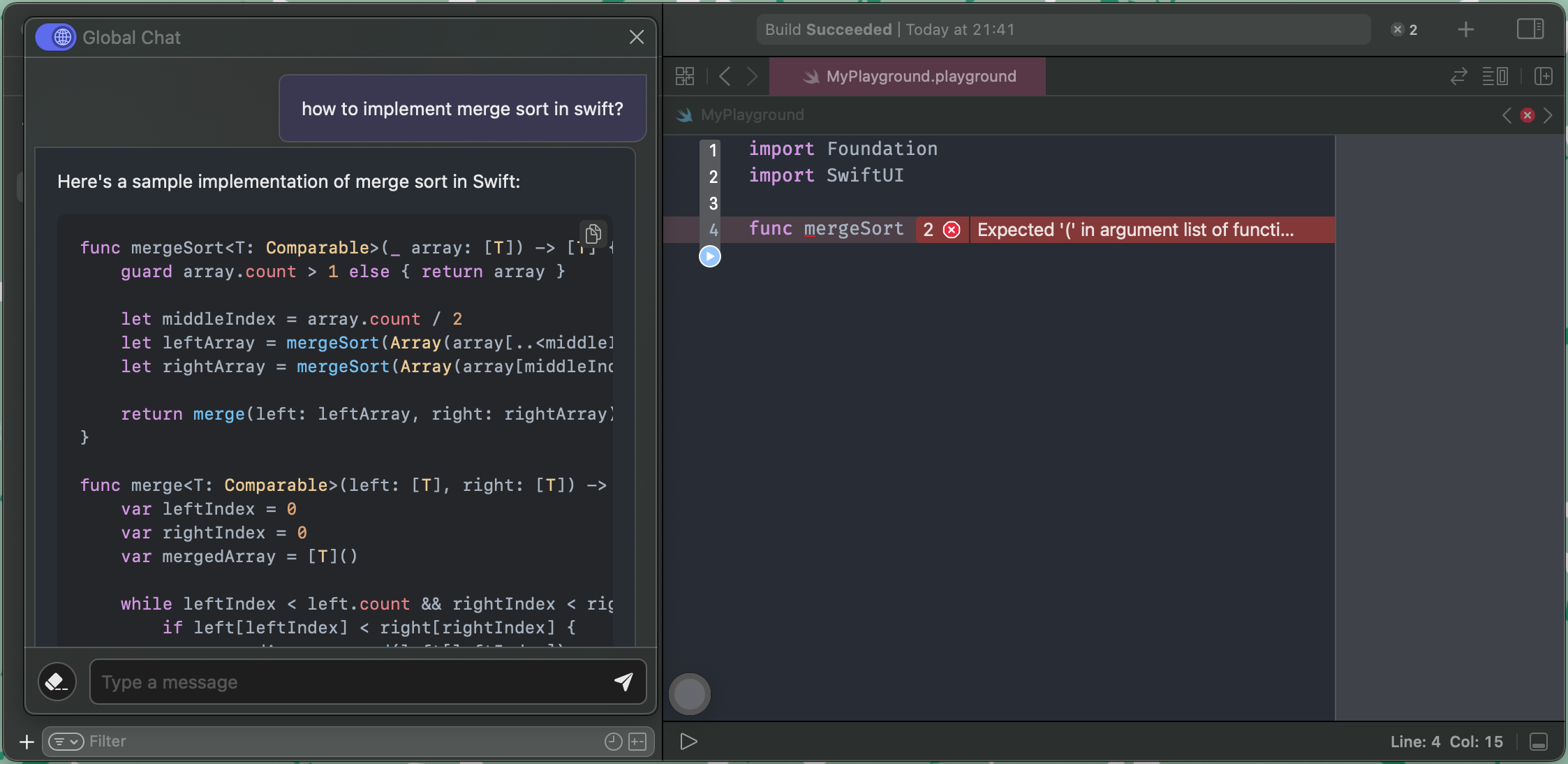}
		\captionsetup[subfigure]{justification=centering}
		\caption{Copilot for Xcode displaying the chat~panel.}
		\label{fig:copilot_for_xcode_screenshot_2}
	\end{subfigure}
	\caption{{\color{black}Interface of Copilot for Xcode integrated with Apple Xcode}. (\textbf{a},{\textbf{b}}) are the actual user interface tool, where a developer can interact with the GitHub Copilot inside the~Xcode.}
	\label{fig:copilot_for_xcode_screenshot}
\end{figure}

With these workarounds, Copilot for Xcode successfully enables Xcode to support GitHub Copilot, as~shown in Figure~\ref{fig:copilot_for_xcode_screenshot}. In~addition, it facilitates the integration of an external chat panel that can access and read the user's code. This chat panel serves as a connection point to leverage LLMs for functionalities such as code explanation and mutation using natural language. The~chat panel can also be extended with plugins to offer additional features, including support for natural language terminal commands. The incorporation of Copilot into Xcode signifies a notable advancement in AI-powered programming for iOS/macOS, expanding the capabilities of language models to widely-used mobile software development tools.

\section{Conclusions}\label{sec:conc}
{\color{black}
This review paper explores the applications of LLMs in software naturalness to gain a better understanding of software development processes and develop applications that cater to the human aspects of software development. Firstly, it provides a background on Big Code and software naturalness, covering topics such as available datasets, tokenization processes, existing language models, and~entropy-based measurements. Secondly, it summarizes recent applications of LLMs trained with Big Code in various tasks, including code generation, code completion, code translation, code refinement, code summarization, defect detection, and~clone detection. Lastly, it discusses the potential challenges and opportunities associated with LLMs in the context of AI-assisted programming~tasks.

Analyzing Big Code repositories and identifying patterns of naturalness can lead to more effective methods for AI-assisted programming. This can ultimately improve the quality and productivity of AI-assisted programming, making it easier for programmers to create high-quality software with fewer errors in less time. In~addition to the challenges faced by LLMs for codes mentioned in this review paper, there are significant opportunities for future work in the field. These opportunities include exploring the development of LLMs that prioritize transparency and interpretability, enabling clearer explanations for code suggestions and bug fixing. Emphasizing the design of AI-assisted programming applications that prioritize fairness, transparency, and~privacy is crucial, as~current research tends to focus primarily on performance and efficiency. By~pursuing these avenues, AI-assisted programming applications can be advanced to be more user-centric, ethically responsible, and~adaptable, ultimately leading to more efficient and effective \mbox{programming workflows.}}

\section*{Acknowledgement}
This work is supported in part by the Ministry of Education, Singapore, under its Academic Research Fund (No. 022307 and AcRF RG91/22) and Google Faculty Award.
\clearpage
\bibliography{reference.bib}

\end{document}